\documentclass[12pt,a4paper]{article}
\usepackage{amsfonts,latexsym}
\usepackage{amsmath,amssymb}
\usepackage{graphicx,color}

\renewcommand{\thefootnote}{\fnsymbol{footnote}}

\begin{document}

\vspace{12mm}

\begin{center}
{{{\Large {\bf Unstable  Schwarzschild-Tangherlini
  black holes \\ in  fourth-order  gravity }}}}\\[10mm]

{Yun Soo Myung\footnote{e-mail address: ysmyung@inje.ac.kr}}\\[8mm]

{Institute of Basic Sciences and Department  of Computer Simulation, Inje University Gimhae 621-749, Korea\\[0pt]}

\end{center}
\vspace{2mm}

\begin{abstract}
We  study  the  stability of  Schwarzschild-Tangherlini (ST) black
holes in fourth-order  gravity which provides a higher dimensional
linearized massive equation.  The linearized Ricci tensor
perturbations are employed to exhibit unstable modes featuring the
Gregory-Laflamme (GL) instability of higher dimensional black
strings, in comparison to the stable ST black holes in Einstein
gravity. It turns out that the GL instability of the ST black holes
in the fourth-order gravity originates from the massiveness, but not
a nature of fourth-order derivative theories giving ghost states.
\end{abstract}
\vspace{5mm}



\vspace{1.5cm}

\hspace{11.5cm}{Typeset Using \LaTeX}
\newpage
\renewcommand{\thefootnote}{\arabic{footnote}}
\setcounter{footnote}{0}


\section{Introduction}

Babichev and Fabbri have shown that the Schwarzschild black holes in
the dRGT massive gravity~\cite{deRham:2010ik,Hassan:2011zd} do not
exist~\cite{Babichev:2013una}.  This was done mainly  by comparing
the linearized massive equation with the four-dimensional linearized
equation around a five-dimensional black string which indicates the
Gregory-Laflamme (GL) instability of $l=0$
mode~\cite{Gregory:1993vy}. In addition, the
authors~\cite{Brito:2013wya} have confirmed this result by
considering the Schwarzschild-de Sitter black hole and extending
$l=0$ mode to generic modes of $l\not=0$.

On the other hand, it is well known that the fourth-order gravity
provides a massive gravity with ghosts~\cite{stelle,Barth:1983hb}.
We have recently shown that
 the Schwarzschild black hole in fourth-order gravity with $\alpha=-3\beta$ (Einstein-Weyl gravity) is unstable
against the linearized-Ricci tensor
perturbation~\cite{Myung:2013doa}. This was shown by comparing the
linearized massive equation for Ricci tensor  with the
metric-perturbation equation around the five-dimensional black
string.  Furthermore, we have studied   the  stability of
Schwarzschild-AdS black hole in
 Einstein-Weyl  gravity which was known to be stable against the metric perturbations~\cite{Liu:2011kf}.
It turned out that solving the linearized-Einstein tensor equation
exhibits unstable modes featuring the GL instability of a
five-dimensional AdS black string~\cite{Myung:2013nna}. These
results are meaningful  because they ensure  that the instability of
the black hole in the Einstein-Weyl gravity is due to the
massiveness, but not a feature  of fourth-order derivative theory
giving ghost states. Also, the mechanism of GL instability plays the
important role in testing the stability of a black hole in a massive
gravity. This is clearly the case that the GL instability could be
mapped to unstable modes for  a black hole in  massive  gravity
theories~\cite{Harmark:2007md}. Importantly, applying  the GL
instability to black holes in fourth-order gravity changes the
stability of the black hole~\cite{Whitt:1985ki,Liu:2011kf}
drastically into the instability~\cite{Myung:2013doa,Myung:2013nna}.

In this work, we wish to investigate the stability of the
Schwarzschild-Tangherlini (ST) black holes (higher dimensional
Schwarzschild black holes)~\cite{Tangherlini:1963bw} in fourth-order
gravity  with $\alpha=\frac{4(1-D)}{D}\beta$. This will be based on
the GL instability of  higher dimensional black strings. It is known
that the ST black holes are dynamically stable against all metric
perturbations of scalar, vector, and tensor in Einstein
gravity~\cite{Ishibashi:2003ap}. The unstable mode of metric
perturbations for black string whose intersection is the ST black
hole, is only the $l=0$ mode of scalar
perturbation~\cite{Kudoh:2006bp}. If the ST black holes are unstable
against the linearized-Ricci tensor perturbation, it will confirm
that
 the instability of
the black hole in fourth-order gravity is due to the massiveness,
but not a nature of fourth-order derivative theory giving ghost
states.
\section{Linearized fourth-order gravity}
We start with the fourth-order gravity action~\cite{Liu:2011kf}
\begin{eqnarray}
S_{\rm FO}=\frac{1}{2\kappa^2}\int d^D x\sqrt{-g} \Big[R +\alpha
R_{\mu\nu}R^{\mu\nu}+\beta R^2\Big] \label{Action}
\end{eqnarray} with two parameters $\alpha$ and $\beta$.
Here the  Gauss-Bonnet term is excluded because (\ref{Action})
admits solutions of the Einstein gravity including the ST black
holes.  From (\ref{Action}), the Einstein equation is derived to be
\begin{equation} \label{equa1}
G_{\mu\nu}+E_{\mu\nu}=0,
\end{equation}
where the Einstein tensor  is given by \begin{equation}
G_{\mu\nu}=R_{\mu\nu}-\frac{1}{2} Rg_{\mu\nu}
\end{equation}
and $E_{\mu\nu}$  takes the form
\begin{eqnarray} \label{equa2}
E_{\mu\nu}&=& 2\alpha
\Big(R_{\mu\rho\nu\sigma}R^{\rho\sigma}-\frac{1}{4}
R^{\rho\sigma}R_{\rho\sigma}g_{\mu\nu}\Big)+2\beta
R\Big(R_{\mu\nu}-\frac{1}{4} Rg_{\mu\nu}\Big) \nonumber \\
&+&
\alpha\Big(\nabla^2R_{\mu\nu}+\frac{1}{2}\nabla^2Rg_{\mu\nu}-\nabla_\mu\nabla_\nu
R\Big) +2\beta\Big(g_{\mu\nu} \nabla^2R-\nabla_\mu \nabla_\nu
R\Big).
\end{eqnarray}
Eq.(\ref{equa1}) allows  a $D$-dimensional ST black hole
solution~\cite{Tangherlini:1963bw}
\begin{equation} \label{sch} ds^2_{\rm ST}=\bar{g}_{\mu\nu}dx^\mu
dx^\nu=V(r)dt^2+\frac{dr^2}{V(r)}+r^2d\Omega^2_{D-2}
\end{equation}
with the metric function \begin{equation} \label{num}
V(r)=1-\Big(\frac{r_0}{r}\Big)^{D-3}.
\end{equation} Hereafter we denote the background quantities with
the ``overbar''. In this case, the background spacetimes is given by
the Ricci-flat  as
\begin{equation}
\bar{R}_{\mu\nu}=0,~~\bar{R}_{\rho\mu\sigma\nu}\not=0.
\end{equation}

We usually introduce the metric perturbation around the ST black
hole to perform the stability analysis
\begin{eqnarray} \label{m-p}
g_{\mu\nu}=\bar{g}_{\mu\nu}+h_{\mu\nu}.
\end{eqnarray}
Then, the linearized Einstein equation is given by
\begin{eqnarray} \label{lin-eq}
\delta G _{\mu\nu}+\alpha\Big[\bar{\nabla}^2\delta
G_{\mu\nu}+2\bar{R}_{\rho\mu\sigma\nu}\delta G^{\rho\sigma}\Big]
+(\alpha+2\beta)\Big[-\bar{\nabla}_\mu\bar{\nabla}_\nu+\bar{g}_{\mu\nu}\bar{\nabla}^2
\Big] \delta R=0,
\end{eqnarray}
where the linearized Einstein tensor, Ricci tensor, and Ricci scalar
are given by
\begin{eqnarray}
\delta G_{\mu\nu}&=&\delta R_{\mu\nu}-\frac{1}{2} \delta
R\bar{g}_{\mu\nu},
\label{ein-t} \\
\delta
R_{\mu\nu}&=&\frac{1}{2}\Big(\bar{\nabla}^{\rho}\bar{\nabla}_{\mu}h_{\nu\rho}+
\bar{\nabla}^{\rho}\bar{\nabla}_{\nu}h_{\mu\rho}-\bar{\nabla}^2h_{\mu\nu}-\bar{\nabla}_{\mu}
\bar{\nabla}_{\nu}h\Big), \label{ricc-t} \\
\delta R&=& \bar{\nabla}^\mu \bar{\nabla}^\nu
h_{\mu\nu}-\bar{\nabla}^2 h\label{Ricc-s}.
\end{eqnarray}
with $h=h^\rho~_\rho$. It is not easy to solve the linearized
equation (\ref{lin-eq}) directly because it is a coupled
second-order equation for $\delta G_{\mu\nu}$ and $\delta R$. Thus,
it would be better  to decouple $\delta R$ from (\ref{lin-eq}).
Taking the trace of (\ref{lin-eq}) leads to
\begin{equation}
\Big[\Big(D\alpha+4(D-1)\beta\Big)\bar{\nabla}^2-(D-2)\Big]\delta
R=0,
\end{equation}
which indicates  that the $D$-dimensional D'Alembertian operator
disappears if one chooses
\begin{equation} \label{i-cond}\alpha=\frac{4(1-D)}{D}\beta.
\end{equation}
 In this
case, the linearized Ricci scalar is constrained to vanish
\begin{equation}
\delta R=0 \label{riccz}
\end{equation}
which implies that  $\delta G_{\mu\nu}\to \delta R_{\mu\nu}$.
Substituting  this into (\ref{lin-eq}) leads to the equation for the
linearized Ricci tensor only
\begin{equation} \label{slin-eq}
\Big[\bar{\nabla}^2-\frac{D}{4(D-1)\beta}\Big]\delta R_{\mu\nu}+
2\bar{R}_{\rho\mu\sigma\nu}\delta R^{\rho\sigma}=0
\end{equation}
If we do not require the condition of (\ref{i-cond}) which
eliminates a massive spin-0 (scalar graviton), we could not go a
further process. In all dimensions, (\ref{i-cond}) enables us to
write the Lagrangian together with an auxiliary field in the
Fierz-Pauli form~\cite{Bergshoeff:2011ri}. For $D=3$, it is a new
massive gravity~\cite{Bergshoeff:2009hq} and the case of $D=4$
corresponds to the Einstein-Weyl gravity~\cite{Lu:2011zk}. For
$D>4$, it gives us a critical gravity on AdS$_{D}$
spacetimes~\cite{Deser:2011xc}.

 After choosing the transverse-traceless
gauge (TTG) \begin{equation}\label{ttg}\bar{\nabla}^\mu
h_{\mu\nu}=0~{\rm and}~h=0,
\end{equation}
the linearized Ricci tensor takes the form
\begin{equation}
\delta R_{\mu\nu}=\frac{1}{2} \Delta h_{\mu\nu}
\end{equation}
with the Lichnerowicz operator \begin{equation} \Delta
h_{\mu\nu}=-\bar{\nabla}^2h_{\mu\nu}-2\bar{R}_{\rho\mu\sigma\nu}h^{\rho\sigma}.
\end{equation}
 Eq. (\ref{slin-eq}) could be expressed as
 a fourth-order differential equation~\cite{Liu:2011kf}
\begin{equation} \label{four-eqq}
\Delta(\Delta+M_D^2)h_{\mu\nu}=0
\end{equation}
which may imply a massless spin-2 equation
\begin{equation} \label{se1-eq}
\Delta h^{m}_{\mu\nu}=0
\end{equation}
 and a massive spin-2 equation
 \begin{equation} \label{se2-eq}
(\Delta+M^2_D)h^{M}_{\mu\nu}=0
\end{equation}
with the $D$-dimensional  mass
\begin{equation}
M^2_D=\frac{D}{4(D-1)\beta}=-\frac{1}{\alpha}.
\end{equation}

On the background of  the ST black hole (\ref{sch}), we rewrite Eq.
(\ref{slin-eq}) as a second-order equation for the linearized Ricci
tensor
\begin{equation} \label{se1m-eq}
\bar{\nabla}^2\delta R_{\mu\nu}+2\bar{R}_{\rho\mu\sigma\nu}\delta
R^{\rho\sigma}=M^2_D \delta R_{\mu\nu}.
\end{equation}
 Similarly, Eq. (\ref{se1-eq}) is expressed as a linearized massless
 equation
 \begin{equation} \label{lem1-eq}
\bar{\nabla}^2h^m_{\mu\nu}+2\bar{R}_{\rho\mu\sigma\nu}h^{m\rho\sigma}=0.
\end{equation}
and  Eq. (\ref{se2-eq}) takes the  linearized massive equation
\begin{equation} \label{lem2-eq}
\bar{\nabla}^2h^M_{\mu\nu}+2\bar{R}_{\rho\mu\sigma\nu}h^{M\rho\sigma}=M^2_D
h^M_{\mu\nu}.
\end{equation}
At this stage, we wish to point out a difference between
(\ref{se1m-eq}) and (\ref{lem2-eq}). The former equation is a
second-order  equation for the linearized Ricci tensor, whereas the
latter is a suggesting second-order equation from the fourth-order
equation (\ref{four-eqq}) for the metric perturbation. It is  known
that the introduction of fourth-order derivative terms give rise to
ghost-like massive graviton~\cite{stelle}, which may automatically
imply instability of a black hole even if a solution
exists~\cite{Myung:2013nna}. Hence, even though
(\ref{se1-eq})[(\ref{se2-eq})] were  used as a linearized massless
[massive] equation on the background of Schwarzschild black
hole~\cite{Liu:2011kf}, their validity is not yet proved because
they seem to be  free from ghost. Splitting (\ref{four-eqq}) into
two second-order equations (\ref{se1-eq})[(\ref{lem1-eq})] and
(\ref{se2-eq})[(\ref{lem2-eq})] is dangerous because the `$-$' sign
in the front of (\ref{se2-eq})[(\ref{lem2-eq})] disappears.  A ghost
state arises from this sign.   Because of a  missing of $-$,  one
may argue that Eq.(\ref{se2-eq})[(\ref{lem2-eq})] by itself does not
represent a correctly linearized equation for studying the stability
of the black hole in the fourth-order  gravity.   However, the
overall $-$ sign in (\ref{se2-eq})[(\ref{lem2-eq})] does not make
any difference unless an external source will be  put on the
right-hand side of (\ref{four-eqq}). Hence, the fourth-order gravity
does not automatically imply the instability of the black hole even
if one uses (\ref{lem2-eq}). Importantly, if one uses
(\ref{se1m-eq}) instead of (\ref{lem2-eq}), one might avoid the
ghost issue  because (\ref{se1m-eq}) is a genuine second-order
equation. This is the reason  why we will take the Ricci tensor
perturbation.

\section{ Instability of ST Black holes
} Let us briefly review the GL stability analysis of
$(D+1)$-dimensional black strings. This can be represented by a
matrix form
\begin{equation}
\left(
  \begin{array}{cc}
   h^{D}_{\mu\nu} & h_{\mu z} \\
  h_{z\nu} & h_{zz} \\
  \end{array}
\right)
\end{equation}
around the $(D+1)$-dimensional black string~\cite{Gregory:1993vy}
\begin{equation}
ds^2_{\rm BS}=ds^2_{\rm ST}+dz^2
\end{equation}
with  the ST element of $ds^2_{\rm ST}$ in (\ref{sch}).

Choosing $h_{\mu z}=h_{zz}=0$,  a $D$-dimensional  $s(l=0)$-mode
takes the form
\begin{eqnarray}
h^{D\mu\nu}=e^{\Omega t}e^{ikz} \left(
\begin{array}{ccccc}
H^{tt}(r) & H^{tr}(r) & 0 & 0  &\ldots \cr H^{tr}(r) & H^{rr}(r) & 0
& 0 &\ldots \cr 0 & 0 &  K(r) & 0 &\ldots \cr 0 & 0 & 0 & \frac{
K(r)}{\sin^2\theta} & \ldots \cr \vdots & \vdots & \vdots& \vdots
&\ddots
\end{array}
\right). \label{evenp}
\end{eqnarray}
The metric perturbation $h^{D}_{\mu\nu}$ satisfies the massive
spin-2 equation
\begin{equation} \label{kaluza-eq}
\bar{\nabla}^2h^{D}_{\mu\nu}+2\bar{R}_{\rho\mu\sigma\nu}h^{D\rho\sigma}=k^2
h^{D}_{\mu\nu}
\end{equation}
together with the TTG of $\bar{\nabla}^\mu h^{D}_{\mu\nu}=0 $ and
$h^{D}=0$. For $k^2\not=0$, eliminating all but $H^{tr}$, Eq.
(\ref{kaluza-eq}) reduces to a second-order equation for $H^{tr}$
\begin{equation} \label{second-eq} A(r;r_0,D,\Omega^2,k^2)
\frac{d^2}{dr^2}H^{tr} +B\frac{d}{dr}H^{tr}+CH^{tr}=0.
\end{equation}
The two boundary conditions are required: a normalizable  solution
at infinity is
\begin{equation}
H^{tr}_{\infty} \sim e^{-\sqrt{\Omega^2+k^2} r},
\end{equation}
while the solution near  the horizon behaves as
\begin{equation}
H^{tr}_{r_0}\sim \frac{1}{(r-r_0)^{1-\frac{r_0\Omega}{(D-3)}}}.
\end{equation}
This is a one-parameter shooting problem with a shooting parameter
$\Omega>0$~\cite{Kudoh:2006bp}. Solving this problem numerically to
search for $\Omega$ and $k$ shows unstable modes for each
$D$-dimensions (see Fig.1 in Ref.\cite{Gregory:1993vy}). Especially
for $e^{\frac{\Omega}{r_0} t} e^{i\frac{k}{r_0} z}$ setting, there
exists a critical non-zero wave number $k_c$ where for
$k<k_c(k>k_c)$, the black string is unstable (stable) against the
metric perturbations.   There is an unstable (stable) mode for any
wavelength larger (smaller) than the critical wavelength
$\lambda_{\rm GL}=2\pi r_0/k_c$: $\lambda>\lambda_{\rm
GL}(\lambda<\lambda_{\rm GL})$. The critical wave number $k_c$
depending on $D$-dimensions is given by~\cite{Harmark:2007md}
\begin{equation} \label{cwn}
\left(
\begin{array}{c|ccccccccccc}
 D & 4 & 5 & 6 & 7 & 8 & 9 & 10 & 11& 12 & 13 & 14 \cr \hline
      k_c (=r_0M_D^c)& 0.88 & 1.24 & 1.60 & 1.86 & 2.08 & 2.30& 2.50 & 2.69 & 2.87 & 3.03 &
      3.18 \cr
\end{array}
\right).
\end{equation}

 For a massive gravity theory
in the Minkowski background, there is correspondence between
linearized Ricci tensor $\delta R_{\mu\nu}$ (\ref{ricc-t}) and Ricci
spinor $\Phi_{ABCD}$ when using the Newman-Penrose
formalism~\cite{Newman:1961qr}.  Here the null real tetrad is
necessary to specify polarization modes of massive graviton, as the
 massive gravity  requires null complex tetrad to
specify six polarization modes~\cite{Eardley:1974nw,Moon:2011gg}.
This implies that in fourth-order gravity with
$\alpha=\frac{4(1-D)}{D}\beta$, one may take linearized Ricci tensor
$\delta R_{\mu\nu}$~\cite{Myung:2013doa}, instead of the metric
perturbation $h_{\mu\nu}$ in Einstein gravity.

 In the above GL
analysis, it is obvious  that the obtained unstable mode is  not a
gauge artifact but a genuine physical mode. This is because imposing
the TTG condition on a symmetric tensor $h^D_{\mu\nu}$ leads to
$D(D+1)/2-(D+1)=(D+1)(D-2)/2$ DOF. Considering the $s$-mode
instability, these $(D+1)(D-2)/2$ DOF reduces to a single DOF of
$H^{tr}$. Up to now, we did not take into account DOF of $\delta
R_{\mu\nu}$ as physical modes. Here we could not choose a gauge
condition like the TTG (\ref{ttg}) directly for a linearized Ricci
tensor $\delta R_{\mu\nu}$. Instead, the linearized version of the
Bianchi identity
\begin{equation}
\bar{\nabla}_{[\mu}\delta R_{\nu\rho]\sigma\kappa}=0 \label{bian}
\end{equation}
implies a relation for $\delta R_{\mu\nu}$ and $\delta R$ when
contracting (\ref{bian}) as
\begin{equation} \label{contbi}
2\bar{\nabla}^\mu \delta R_{\mu\nu}-\bar{\nabla}_\mu \delta R=0,
\end{equation}
leading to the well-known Bianchi identity $\bar{\nabla}^\mu \delta
G_{\mu\nu}=0$ the linearized Einstein tensor. Considering $\delta
R=0$ (\ref{riccz}), the contracted Bianchi identity (\ref{contbi})
reduces  to
\begin{equation} \label{tbian}
\bar{\nabla}^\mu \delta R_{\mu\nu}=0
\end{equation}
which plays a role of the transverse condition. Taking into account
(\ref{tbian}) together with the traceless condition (\ref{riccz})
leads to DOF for $\delta R_{\mu\nu}$ as
\begin{equation}
\frac{D(D+1)}{2}-(D+1)=\frac{(D+1)(D-2)}{2}
\end{equation}
which is exactly  the same DOF for a metric tensor $h^D_{\mu\nu}$.

At this stage, we emphasize again that (\ref{se1m-eq}) is considered
as the second-order equation with respect to $\delta R_{\mu\nu}$,
but not the fourth-order equation (\ref{four-eqq})  for
$h_{\mu\nu}$. Hence, we propose $\delta R_{\mu\nu}$ as physical
observables on the ST black hole background.  Similarly, we find the
same equation (\ref{se1m-eq})  when substituting $h^{D}_{\mu\nu}$
and $k^2$ into $ \delta R_{\mu\nu}$ and $M^2_D$ in
(\ref{kaluza-eq}). Also, we impose (\ref{tbian}) and (\ref{riccz})
to find the $s$-mode instability of $\delta R_{\mu\nu}$.
Accordingly, the relevant equation for a physical mode of $\delta
R^{tr}$ takes the same form
\begin{equation} \label{secondG-eq} A(r;r_0,\Omega^2,M_D^2)
\frac{d^2}{dr^2}\delta R^{tr} +B\frac{d}{dr}\delta R^{tr}+C\delta
R^{tr}=0
\end{equation}
which shows  unstable modes for
\begin{equation} \label{unst-con}
0<M_D<M_D^c=\frac{r_0M^c_D}{r_0}=\frac{k_c}{r_0} \end{equation} with
the $D$-dimensional mass
\begin{equation} M_D=\sqrt{\frac{D}{4(D-1)\beta}}.
\end{equation}
and the $D$-dimensional  critical mass $M_D^c$(\ref{cwn}).
Especially, even if one uses (\ref{lem2-eq}) as a linearized massive
equation~\cite{Liu:2011kf}, our conclusion remains unchanged because
(\ref{lem2-eq}) and (\ref{se1m-eq}) are the same equation for
different tensors. It turned out that for $M^2_D=0$, the ST black
holes are stable against the metric perturbations in Einstein
gravity when using (\ref{lem1-eq})~\cite{Ishibashi:2003ap}.

Consequently,  the instability arises  from the massiveness
($M_D>0$) but not from a feature of the fourth-order equation which
gives the $-$ sign (ghost=negative norm state) when splitting it
into two second-order equations.  This means  that the ST black
holes in fourth-order gravity  with $\alpha=\frac{4(1-D)}{D}\beta$
do not exist and/or they do not form in the gravitational collapse.

Finally, we could not carry out the stability of the ST black holes
in fourth-order gravity with arbitrary $\alpha$ and $\beta$ because
the linearized equation (\ref{lin-eq}) is a coupled equation for
$\delta G_{\mu\nu}$ and $\delta R$, leading to a fourth-order
equation for $\delta \tilde{R}_{\mu\nu}=\delta R_{\mu\nu}-\delta R
\bar{g}_{\mu\nu}/4$~\cite{Barth:1983hb}.

 \vspace{1cm}

{\bf Acknowledgments}
 \vspace{1cm}

 This work
was supported by the National Research Foundation of Korea (NRF)
grant funded by the Korea government (MSIP)
(No.2012-R1A1A2A10040499) and  by the Korea government (MSIP)
through the Center for Quantum Spacetime (CQUeST) of Sogang
University with grant number 2005-0049409.

\end{document}